\documentclass[twocolumn,preprintnumbers,amsmath,amssymb,prb]{revtex4}
\usepackage{epsfig,graphicx}

\usepackage{dcolumn}
\usepackage{bm}

\begin{document}

\title{Thermodynamic and transport anomalies near isotropic-nematic
phase transition}

\author{Prasanth P. Jose and Biman Bagchi}
\email{bbagchi@sscu.iisc.ernet.in}
\affiliation{Solid State and Structural chemistry Unit,
Indian Institute of Science,
Bangalore - 560012, India.}

\date{\today}

\begin{abstract}

A theoretical study of the variation of thermodynamic
and transport properties of
calamitic liquid crystals across the isotropic-nematic phase transition
is carried out by calculating the {\it wavenumber (k) and time (t)} dependent
intermediate scattering function of the liquid, via computer simulations of
model nematogens. The objective is to understand the experimentally observed anomalies 
and sharp variation in many thermodynamic and transport properties, namely specific heat $C$,
sound attenuation coefficient $\Gamma$, thermal diffusivity $D_T$
and sound velocity $c_s$ are as the I-N transition is approached from the isotropic side.
The small wavelength limit of the calculated intermediate scattering function
$F(k,t)$ is used to obtain the ratio of specific heats $\gamma$ and
other properties mentioned above. We find that all of them show
non-monotonic variations near the I-N transition, with $\Gamma$
showing a cusp-like behavior.  We suggest that the observed
anomalous features are a direct consequence of the existence of pseudo-nematic
domains in the system near the phase boundary and the melting and formation
of such domains give rise to sound attenuation and also to the observed specific heat
anomaly. A theoretical description of these anomalies should invoke 
translation-rotation coupling at molecular level. While the heterogenous dynamics 
observed here bear resemblance to that in deeply supercooled liquids near glass transition, 
the thermodynamic anomalies articulated here are largely absent in supercooled liquids.
\end{abstract}
\maketitle

\section{Introduction}

Time correlation functions of wavenumber dependent collective
number density and orientational fluctuations in a molecular
liquid play important role in understanding many microscopic aspects of dynamics
in molecular liquids. The wavenumber dependent collective
correlation functions of fluctuations bridge the gap between the
single particle relaxation and the relaxation in the hydrodynamic
limit ($k\rightarrow 0$ and $\omega \rightarrow 0$)
\cite{vanhove,zwanzig:tcf}. These correlation functions can be
measured in the experiments such as light scattering and inelastic
neutron scattering \cite{bap,cal,ham,baz,bay,rah3:sk,rah4:sk}. In the
hydrodynamic limit, the expressions of these correlation functions
can be obtained from the linearized Navier-Stokes equations
\cite{bap,cal,ham}. This linearized hydrodynamic model and various molecular
extensions of them  have been useful in interpreting
the results obtained from the computer simulations and the experiments
of simple liquids in equilibrium
\cite{mountain,bap,ham}. The time correlation function analysis of the
liquid state uses well-known expressions for the transport coefficients and
the hydrodynamic behavior.  Such an analysis includes the wavelengths
and frequencies that are comparable to inter atomic distances and
microscopic relaxation rates. Deviations from the hydrodynamic
behavior often appear in this analysis above some critical wavenumber.
The wavenumbers near this critical distance may be referred as
intermediate wavenumbers. Even after many years of investigation
by different researchers, a complete microscopic understanding of
the relaxation of collective fluctuation in liquids
remains incomplete. At intermediate wavenumber,  the correlation
functions of collective fluctuations deviate from their hydrodynamic
behavior; here the distance referred to is comparable to that of
inter particle separation. The collective relaxation in the
intermediate wavenumber regime can be treated in the framework
of extended hydrodynamics \cite{bay,ham,cohen,yip1}.

 In contrast to simple liquids, the studies of molecular liquids are
bound to be complex due to the presence of the internal degrees of
freedom, which couple to the conserved modes and act as additional
relaxation channels for fluctuation. Compared to simple liquids,
there are several parameters of intermolecular potential, like
aspect ratio and geometry of molecule, which may affect the
microscopic relaxation of molecular liquids \cite{mls1,mls2,yip2}.

There are several attempts to study the relaxation
of the hydrodynamic modes in molecular liquids, however,
such an analysis  near the I-N transition is found to be
absent in the literature. It is interesting to study this aspect
of I-N transition, especially in the context of the recent
experimental results by Fayer and coworkers
\cite{fay1,bb:fay1,bb:fay2,fay2,fay3,fay4,fay:nem}, where the orientational
degree of freedom shows relaxation dynamics similar to that found in
supercooled liquids. In another work Li {\it et al.} \cite{fay:nem} presented a
mode coupling theory which could reproduce the results observed in the
experiments. However, they did not provide origin of the
mode which couples to the orientational degrees of freedom.
On the other hand, the experimental study
of collective relaxation of other modes show the
signature of the second order phase transition \cite{tanaka}.
In the dynamic light scattering experiments near the I-N transition,
the spectra of the orientational degrees of freedom
mask dynamics of other modes.  Mode-selective dynamic light scattering
experiments of Takagi {\it et al.} \cite{tanaka} on 5CB and 7CB
liquid crystal samples have studied the dynamics of modes usually
hidden under fluctuation of the orientational mode. They have measured
the thermal diffusion, orientational relaxation and propagating
sound modes of those liquid crystal samples. Using photopyroelectric
technique, Marinelly {\it et. al} \cite{tdiff} have measured the
thermal conductivity and thermal diffusivity of aligned nCB samples.
They have found that the thermal conductivity shows a discontinuity
near the I-N transition, while, the thermal diffusivity shows a dip.
In older studies of  ultrasonic investigations of
Eden {\it et al.} \cite{snd:expt1} on MBBA samples with out any
preferred orientation with emphasis given near transition
temperature $\sim$ 44 $^oC$, they found the acoustic
relaxation time and relaxation strength increases near I-N transition.

It is now worth while to study effect of I-N transition
on conserved degrees of freedom and also on the
thermodynamic and transport properties. In this paper, the study of
the collective transport properties of  the nematogens has been restricted chiefly to
the pre-transition region, where, even though there is an anisotropy
due to the formation of pseudo nematic domains,  they relax slowly
to remove the anisotropy. However, as the orientational relaxation
slows down close to the I-N transition, the relaxation of the
anisotropy also slows down. Since the I-N transition is weakly
first order, there are prolonged pre-transitional effects, which
enable the observation of the effect of orientational ordering
in the structural relaxation for a range of state points across
I-N transition.

Since I-N transition occurs due the formation of the orientational
cage that freezes rotation of the molecules, as the I-N
transition is approached, the exchange of the angular momentum
with the cage gives rise to a strong back scattering region in
the angular velocity auto correlation function (see Ref.
\cite{bb:ppj2}).

It is of course of interest to study what signatures the formation
and melting of pseudo-nematic domains have on the
transport properties, like thermal diffusivity, and also on the
thermodynamic properties like sound velocity and specific heat.
Earlier we presented a study of the variation of viscosity which shows weak
anomaly across the isotropic-nematic phase transition. The present study
is a continuation of our earlier study \cite{bb:ppj2} but now extended
to include many thermodynamic and transport properties. Just as in the
case of viscosity, it is not clear in the beginning what role formation
of orientational order can have on properties which are largely
determined by translational degrees of freedom.

Investigations reported in this paper are based on
the simulation of a model molecular liquid with aspect
ratio $\sim 3$ and can be approximated by prolate ellipsoids with two
of the minor axis are equal.  As reported earlier
\cite{bb:ppj1,bb:ppj2},  this system undergoes an I-N transition as
the density increases from $\rho^*$ =0.285 to  $\rho^*$ =0.315,
gradually along an isotherm at $T^*$ = 1.0. The  single particle
relaxation dynamics in this system shows usual behavior as that of an
ordinary liquid, except the presence of a marked anisotropy near
I- N transition.

The present study on the relaxation of fluctuation of
collective density modes focuses on the relaxation
dynamics at different wavenumbers.  The
intermediate scattering function calculated from the simulations is
fitted to the linearized hydrodynamic model of the intermediate
scattering function in order to find the thermodynamic and transport
properties of the system. Thermodynamic and transport
properties obtained from the fitting parameters show  non-monotonic
behavior at different state points across the I-N transition.
The sound velocity obtained from the peak position of the Brillouin
peak \cite{ham,bay,baz} shows that the adiabatic sound velocity
increases across the I-N transition; except a deviation from it's
monotonic increase near the transition.

An important (and we believe) new result of the present study is
the appearance of anisotropy in the wavenumber dependence of
the dynamic structure factor, F(k,t). The latter becomes a function
of the direction of propagation of wavevector, ${\bf k}$.

The remaining sections of this paper are organized as follows.
Section \ref{ds2} discusses the definition of the time
correlation functions and their expressions in the
hydrodynamic limit, which are used in the analysis.
Details of simulations are presented in the section \ref{sim1}.
The section \ref{ds3} presents the results and discussions
on the time correlation functions calculated from the
simulations. The summary of the work is presented in the
section \ref{ds4}.

\section{\label{ds2} Collective correlation functions of density}
For a system that consists of N linear molecules there are 3N
coordinates of position and 2N coordinates of orientation.
In this system, density is a function the
positions as well as orientations $\rho=\rho({{\bf r},{\bf\Omega}})$.
The number density of the system  $\rho({\bf r})$ can be obtained from
$\rho({\bf r},{\bf \Omega})$ by averaging over the orientations
${\bf \Omega}$. Note that, the orientation of the molecule is not
a conserved quantity.

The intermediate structure factor $F(\bf k,\Omega,\Omega^\prime,t)$
which measures the relaxation of  density fluctuations in the
molecular liquid is defined as \cite{bb:ch1},
\begin{equation}
F({\bf k},\Omega,\Omega^\prime,t)=\frac{1}{N}
\langle\rho(-{\bf k},\Omega, 0)
\rho({\bf k},\Omega^\prime, t)\rangle.
\end{equation}
This correlation function also accounts for the correlation between
the orientation and density at two points separated in space
and time. This is a generalization of the
intermediate scattering function \cite{ham,bay,baz} of a
monatomic liquid. On averaging over orientation,
$F({\bf k},\Omega,\Omega^\prime,t)$ gives
$F({\bf k},t)$, which is the intermediate scattering function of
 number density of the system. In the theory of
the linearized hydrodynamics, $F({\bf k},t)$ can be expressed
as \cite{bap,ham}
\begin{eqnarray}
F( k \rightarrow 0,t)&=& S(k \rightarrow 0)
[ a\; e^{-t/\tau_a}+\nonumber \\
&&(1-a)\;e^{-t/\tau_b}\; \nonumber \\ & &
(cos(\omega_0 t)+b\;sin(\omega_0 t))],
\label{fkthyd}
\end{eqnarray}
where $S(k)$ is the static structure factor, $a=(\gamma-1)/\gamma$
($\gamma$ is the ratio of specific heats ($C_p/C_v$)),
$\tau_a\;=\;1/D_Tk^2$ ($D_T$ is the thermal diffusivity),
$\tau_b\;=\;1/\Gamma k^2$ ($\Gamma$ is the sound wave
damping constant), $\omega_0=ck$ ($c$ is the adiabatic
sound velocity) and $b=k((3\Gamma-D_v)/\gamma c)$ ($D_v$
is the kinematic viscosity).

The intermediate scattering function $F(k,t)$ and the structure
factor gives information about the variation of the
thermodynamical as well as transport coefficients. The $S(k)$
at the lowest value of $k$ gives the isothermal
compressibility
\begin{equation}
\chi_T=\rho^{-1}\left(\frac{\partial \rho}{\partial p}\right)_T,
\end{equation}
of the system.
$\chi_T$ of the liquid across the I-N transition obtained from
$S(k)$ by the relation
\begin{equation}
\chi_T=\frac{\beta}{\rho}\lim_{k\rightarrow0}S(k)
\end{equation}

\section{\label{sim1}Details of the simulations}

\begin{figure}
\epsfig{file=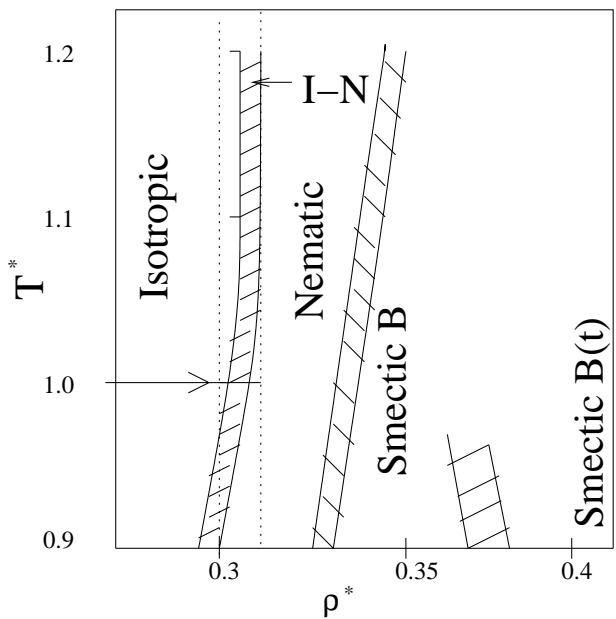,width=8.3cm}
\caption{The phase diagram of the Gay-Berne ellipsoids.
The simulations are carried out at the
temperature and densities indicated by the arrow.}
\label{pdiag}
\end{figure}

Molecular dynamics simulations have been carried out mostly for
a system of 1125 Gay-Berne ~\cite{berne1,berne2} ellipsoids at
temperature $T^*$=1 near I-N transition in a micro-Canonical
ensemble. The form of the modified inter-molecular Gay-Berne
potential used in the simulation is
 ~\cite{cra:acp,bb:rv1,bb:rv2,bb:vas1,bb:ppj1}
\begin{widetext}
\begin{equation}
U=4\epsilon(\hat r,\vec u_i, \vec u_j)\
\left[ \left(\frac{\sigma_s}
{r-\sigma(\hat r,\vec u_i,\vec u_j)+\sigma_s}\right)^{12}
-\left(\frac{\sigma_s}
{r-\sigma(\hat r,\vec u_i,\vec u_j)+\sigma_s}\right)^{6}
\right]
\label{gb1:eq}
\end{equation}
where $\hat r$ is the unit vector that passes through the
center of mass of a pair of molecules, $\vec u_i$ and
$\vec u_j$ unit vectors that passes through
the major axis of a pair of ellipsoidal molecules.
$\epsilon$ and $\sigma$ give the strength and range of
interaction,
\begin{equation}
\sigma(\hat r,\vec u_i,\vec u_j)=\sigma_s
\left[1-{\frac{\chi}{2}}
\left(\frac{(\vec u_i \cdot \hat r+\vec u_j
 \cdot \hat r)^2}{1+\chi(\vec u_i \cdot \vec u_j)}
+\frac{(\vec u_i\cdot\hat r-\vec u_j\cdot\hat r)^2}
{1-\chi(\vec u_i\cdot u_j)}
\right)\right]
\label{gb2:eq}
\end{equation}
$\sigma_s$ is double of the minor axis b, $\kappa$ gives molecular
elongation (aspect ratio), which is the ratio of  end-to-end to
side-to-side diameters, $\kappa=\sigma_e/\sigma_s $.
The aspect ratio of ellipsoids used in this simulation is 3.
Here $\chi$ is defined as $\kappa$ as $\chi=(\kappa^2-1)/(\kappa^2+1)$
\begin{equation}
\epsilon(\hat r,\vec u_i,\vec u_j)=\epsilon_0
\left[1-{\chi^2}(\vec u_i \cdot \vec u_j)
\right]^{-\frac{1}{2}}
\left[1-{\frac{\chi^{\prime}}{2}}
\left(\frac{(\vec u_i \cdot \hat r+\vec u_j
 \cdot \hat r)^2}{1+\chi(\vec u_i \cdot \vec u_j)}
+\frac{(\vec u_i\cdot\hat r-\vec u_j\cdot\hat r)^2}
{1-\chi(\vec u_i\cdot u_j)}
\right)\right]^2
\label{gb3:eq}
\end{equation}
\end{widetext}
where $\epsilon_0$ is the energy parameter and
$\chi^{\prime}=(\sqrt{\kappa^{\prime}}-1)/(\sqrt{\kappa^{\prime}}+1)$
($\kappa^{\prime}=\epsilon_s/\epsilon_e $
gives the strength of interaction which is side-to-side to end-to-end
well depths). The value of $\kappa^{\prime}$ used in the simulation is
5 ~\cite{bb:rv2,deMug2,vis1}. The scaling used
for moment of inertia is $I^*=I/m{\sigma_0}^2$. The density
is scaled  in the simulation as $\rho^*=\rho{\sigma_0}^3 $
and the temperature is scaled as  $T^*=k_bT/\epsilon_0$.
The equation of motion is integrated with reduced time
($t^*=(m\sigma_0^2/\epsilon_0)^{1/2}$) steps with
$\Delta$t = 0.002 $t^*$. The ellipsoid used in the
simulation has minor axis b = 0.5 and major axis a = 1.5
(in reduced units). The simulations are done at the state
points near the pre-transition region  of phase diagram,
shown in {\bf figure ~\ref{pdiag}} The translational and
rotational motions are solved using leap-frog algorithm.
The order parameter changes dramatically in this system after
density increases beyond 0.3. This is in accord with our previous
works on same system ~\cite{bb:ppj1}.

\begin{figure}
\epsfig{file=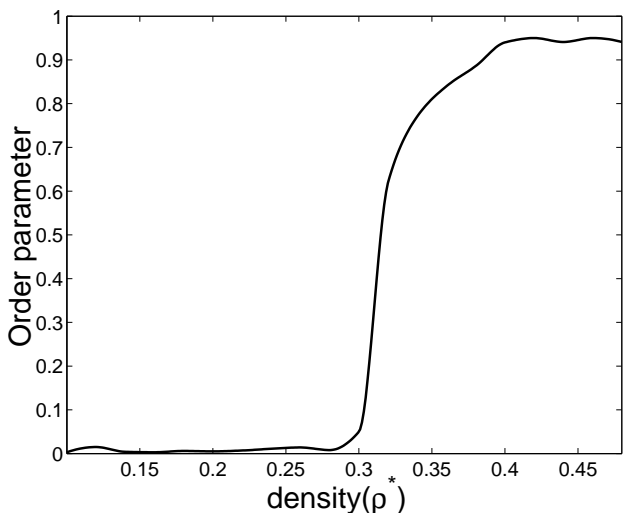,width=8.3cm}
\caption{The variation in the order parameter with density.  }
\label{op}
\end{figure}

The simulation starts from an equilibrated configuration of
ellipsoids. Initial configuration  of the ellipsoids is
generated from  a cubic lattice and then the simulation is
run for two hundred thousand steps to obtain the equilibrium
configuration. During the equilibration steps the temperature
is scaled so that the system is in equilibrium at this
particular temperature.

In  figure ~\ref{pdiag} , we show the phase diagram of
ellipsoids with aspect ratio 3 interacting through Gay-Berne
potential. In the diagram we also show by arrow the densities
studied at temperature ($T^*=1.0$). {\bf Figure ~\ref{op}} shows
the variation of order parameter $S$ ~\cite{bb:ppj1} with density.

The molecular dynamics simulations ran for 10 million steps to
generate trajectories. We have  calculated the collective correlation
functions across I-N transition from this data. The wavevectors
are chosen to be in the $x$, $y$ and $z$ directions of the simulation
box. The difference in the calculated correlation functions
in different directions is a measure of the anisotropy developed
in the system. Relaxation of  wavenumber dependent
correlation functions has been studied at three wavenumbers.
The smallest wavenumber studied is at $k=2\pi/L$, where $L$ is the
length of the simulation box. The lowest value of the wavenumber
$k$ as the density increases from 0.285 to 0.32 with a grid of 0.005
is $\approx$ 0.4. The latter is studied with a grid of $\delta k \approx; $ 0.06.

\section{\label{ds3}Results and discussions}
\begin{figure*}
\epsfig{file=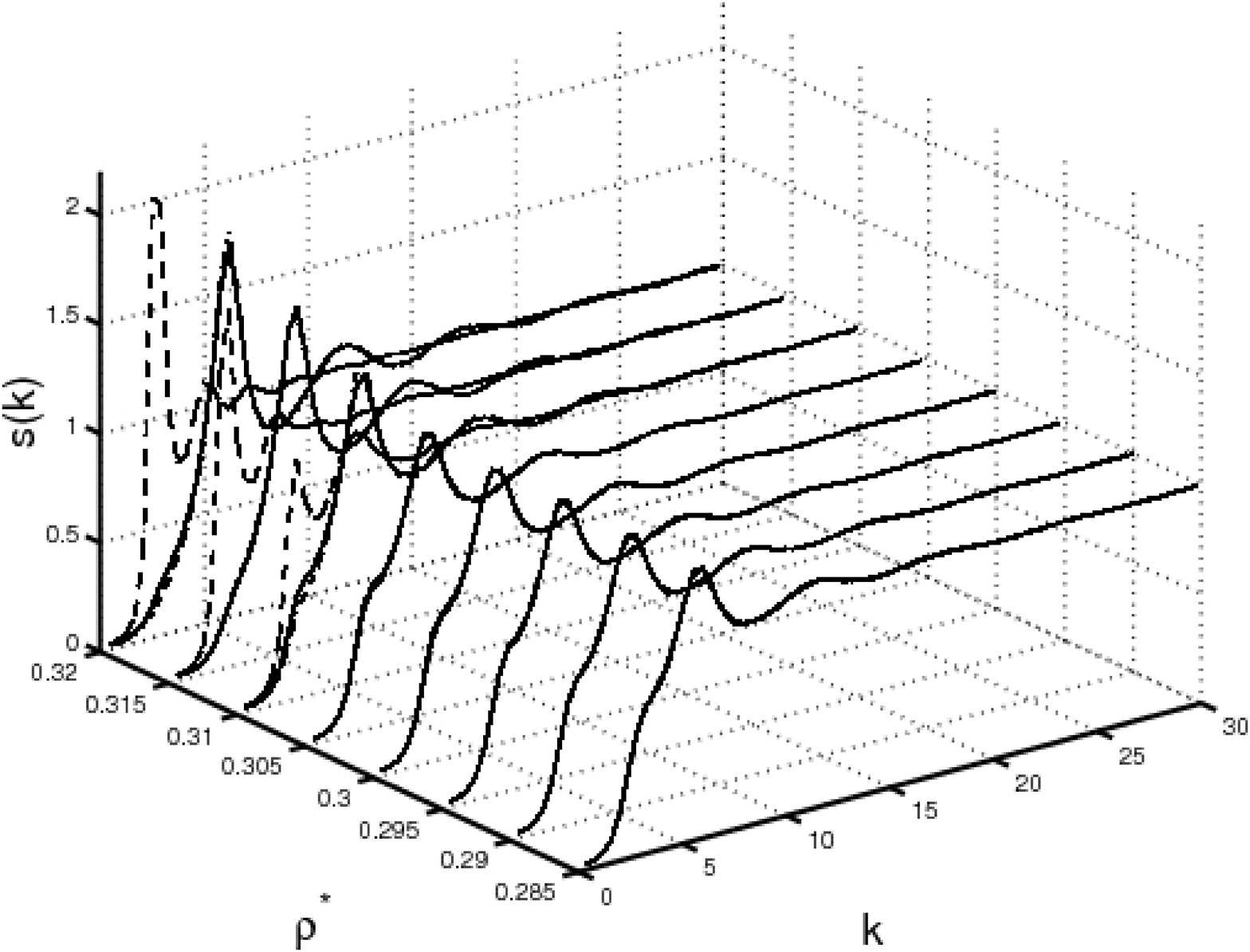,width=11.3cm}
   \caption{The calculated structure factor of the system is
plotted for different densities against wavenumber. The three
 lines at each density shows the structure of the liquid
directions of  $x$, $y$ and $z$: shown respectively
by solid , dashed-dot  and dash  lines. }
   \label{sk}
\end{figure*}
\begin{figure*}
\epsfig{file=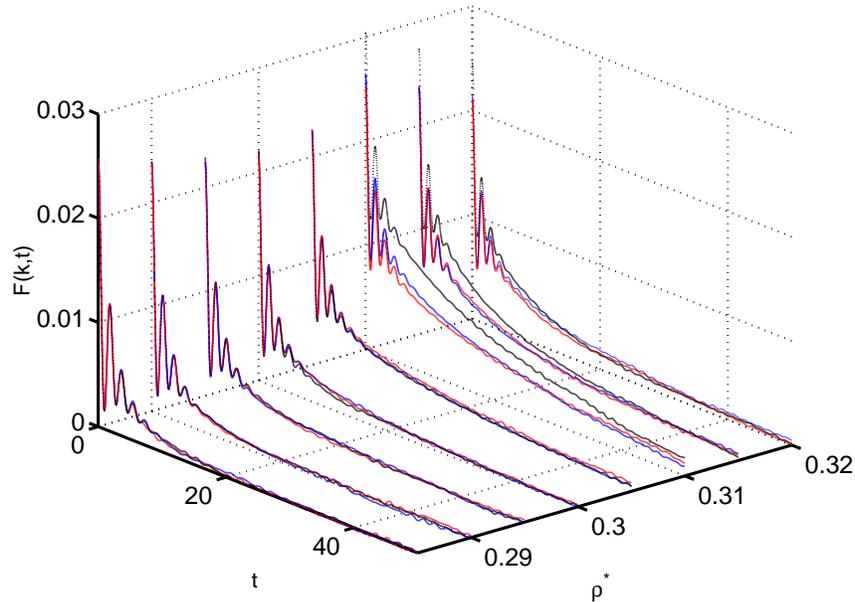,width=11.3cm}
   \caption{The  intermediate  scattering function $F(k,t)$
at lowest k
plotted for different densities against time. The three
 lines at each density shows the structure of the liquid
directions of  $x$, $y$ and $z$: shown respectively
by solid (blue) , dashed-dot (red)  and dash (black)  lines. }
   \label{fkt1}
\end{figure*}

\begin{figure}
\epsfig{file=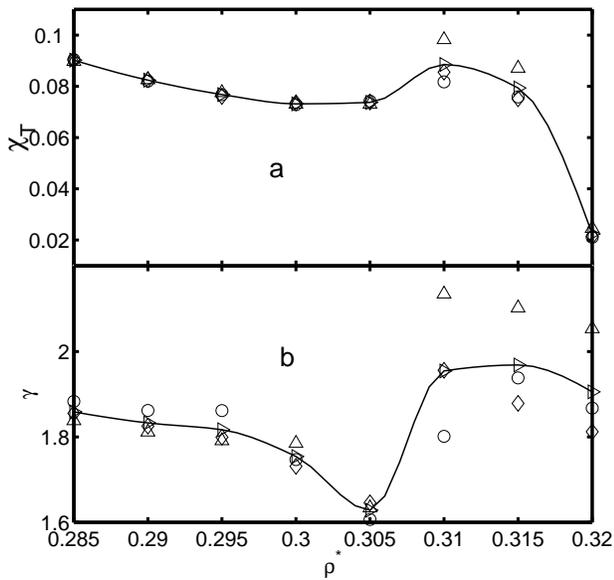,width=8.3cm}
   \caption{The variation of isothermal compressibility ($\chi_T$) (a)
and the ratio of the specific heats ($\gamma$) (b) are plotted against
density across I-N transition  and variation in the isothermal
compressibility of the liquid across the I-N transition. The symbols
diamond, circle, triangle, and left triangle represent respectively
the data obtained from the analysis of dynamic intermediate
scattering function along $x$, $y$, $z$ directions and the average
value of them. }
   \label{hydr_st}
\end{figure}
\begin{figure}
\epsfig{file=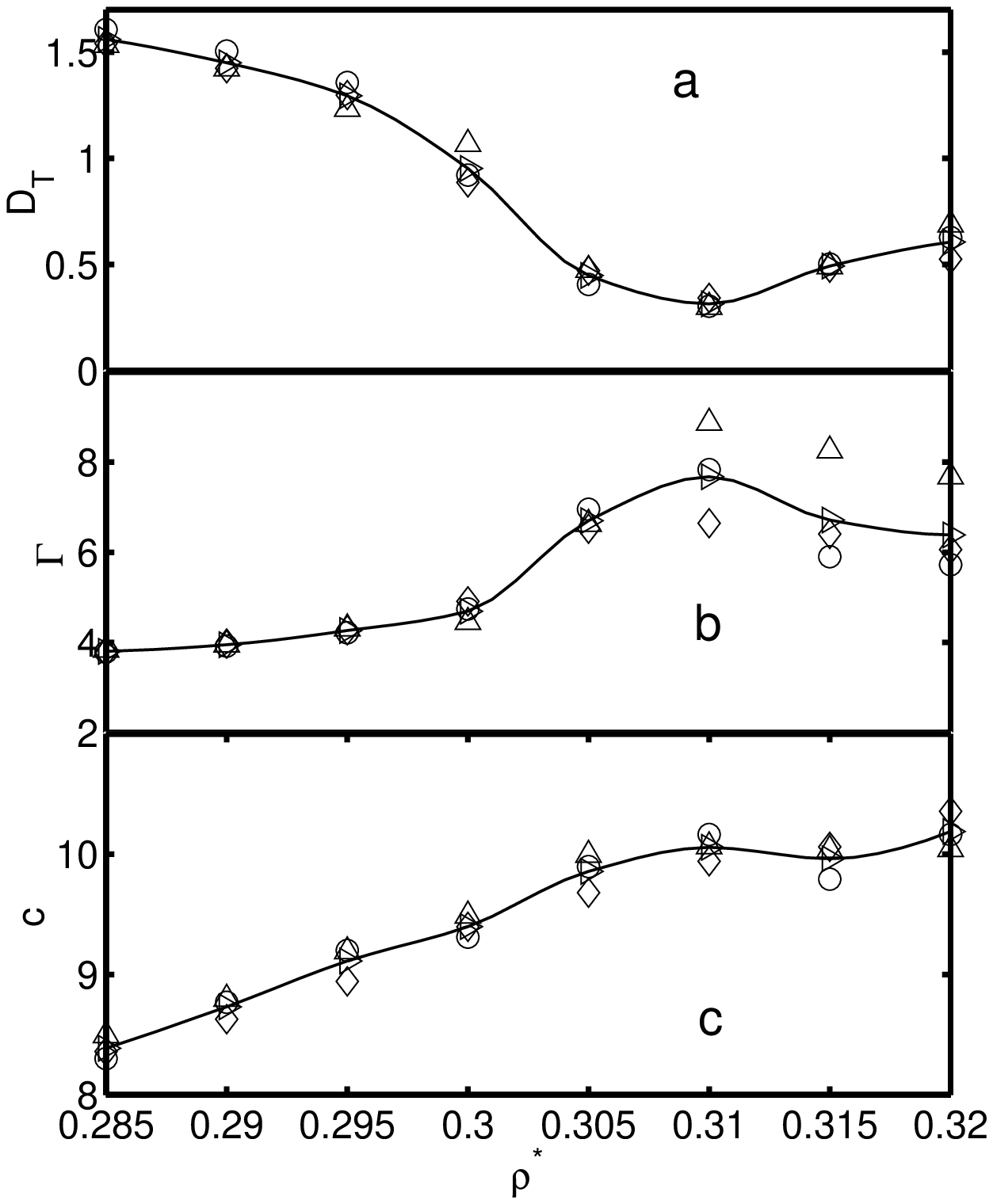,width=8.3cm}
   \caption{Hydrodynamic parameters obtained from the fit of
the intermediate scattering function are plotted against densities
across the I-N transition. (a) shows the variation of the thermal
diffusivity. (b) shows the variation of the sound dispersion
coefficient. (c) shows the variation of the sound velocity.
The symbols diamond, circle, triangle, and left triangle represent
respectively  the data obtained from the analysis of dynamic
intermediate scattering function along $x$, $y$, $z$ directions
and the average value of them.}
   \label{hydr_dy}
\end{figure}
\subsection{\label{ds3a} Density fluctuations}
Figure \ref{sk} shows the variation of the structure factor
$S(k)$ across the I-N transition. As density increases a structure
of the liquid differs at different directions. In one of the
direction (in case of present simulation) the director of the
system is oriented approximately in the $z$ direction. Note that
the time scale of the director relaxation is much slower than the
the density relaxation \cite{bb:ppj1}. This enables the study of the
system in different direction parallel and perpendicular to the
director  of the system.  The development
of the anisotropy in the system is not appreciable at the smallest
wavenumbers. However, {\it at large wavenumbers correlation of density
increases in the $x$ and $y$ directions, at the same time along
the $z$ direction density correlations vanish at larger wavenumbers.}
Therefore, in this system the orientational relaxation is mainly due to
the relaxation of the local orientational cage rather than the global
rotation of pseudo-nematic domain. This enables the study of the
$k_\parallel$ and $k_\perp$ without the use of the use of an external
orienting field. Note that this possible only because the time scale of
the were these correlations measured are short.

Figure ~\ref{fkt1} shows the total intermediate scattering
function for lowest wavenumber chosen for the
study of relaxation, at different state points of the
Gay-Berne liquid  across I-N transition  The $F(k,t)$ at
$k=k_L$ shows hydrodynamic behavior. The rapid oscillation
of the calculated $F(k,t)$ at this wavenumber is related to the
adiabatic sound propagation in the medium according to
linearized hydrodynamics \cite{bap,bay,ham}. As density fluctuation
increases the short wavenumber shows a rapid growth of correlation at the
transition point. This can be an effect of  the
coupling of the between the orientational and the translational modes. This observation
is not the artifact of the development of the global anisotropy of
the system due to nematic domain formation. If that is the case
the growth of correlation of the density fluctuation will be different in the case
of $x$, $y$, and $z$ directions of the simulation box. At this wavevector in all the
three direction density relaxation shows same feature of growth
of correlation near the transition point. Note that highest density
$\rho^*$ = 0.32, which is the nematic phase, where correlation of static
density fluctuation at the lowest wavenumber reduces.

\subsection{Thermodynamic and Transport coefficients}

$F(k,t)$ is fitted to it's hydrodynamic form (Eq. \ref{fkthyd} )
at the lowest value of $k$. From the  parameters of the fit of
$F(k,t)$, at $k=k_L$, various thermodynamic parameters and transport
coefficients can be obtained. Figures \ref{hydr_st} and \ref{hydr_dy} shows
various thermodynamic parameters and transport coefficients
calculated from the fit of $F(k,t)$ for  $k=k_L$. The variation of the
isothermal compressibility of the system at I-N transition is  shown in the
figure  \ref{hydr_st}(a). The isothermal compressibility shows a hump
near I-N transition.  The ratio of the
specific heats $\gamma$ ($\gamma=1/1-a$) is plotted against density is
shown in figure \ref{hydr_st}(b).
The variation of thermal diffusivity
($D_T=1/\tau_a k^2$) against density shown in
figure \ref{hydr_dy}(a), {\it which shows a dip near the I-N
phase boundary.} A similar feature was observed in the experiment
of Marinelly {\it et al.} in the study of
thermal diffusivity of aligned nCB samples \cite{tdiff}.
$\gamma$ shows a rapid increase
at the transition point. Another transport coefficient that can be calculated
from the fitting parameters is the sound wave damping constant, $\Gamma$
($\Gamma=1/\tau_b k^2$). The variation of $\Gamma$ against density is
shown in figure \ref{hydr_dy}(b).  $\Gamma$ shows a rapid growth
 near the I-N transition. This results of   the variation sound damping
constant when I-N transition is approached, is in qualitative agreement
with the results of the Eden {\it et al.}\cite{snd:expt1}.
In figure \ref{hydr_dy}(c), the adiabatic  sound velocity deduced
from the position of the Brillouin peak is plotted against density.
 Except for a few points near the I-N transition,
the adiabatic sound velocity $c$ shows a monotonic increase.
In figure \ref{hydr_dy}(c), sound velocity shows a plateau like
region very near the I-N phase boundary.
A qualitative picture of the effects of the emergence of increased static correlation
on the collective dynamics of
a system of nematogens near the I-N transition can be obtained
from the analysis presented above. In the isotropic phase of the
Gay-Berne liquid, the relaxation of density is isotropic due the
rapid rotation of the molecules.  But, near the I-N transition
due to the freezing of individual rotation of the molecules,
the anisotropy persists for long, which makes the density relaxation
anisotropic. This may be viewed in the perspective of supercooled
liquids, where it take very long time to relax anisotropy.
In supercooled liquids there is no direction for
the anisotropy.  Near the I-N  transition, in the isotropic liquid,
the anisotropy is due to the coupling of number density to the
transient director of the system. In the case of the
supercooled liquids the density relaxation also slows down several
orders of magnitude \cite{got1:rw,got2:rw}. However, in the Gay-Berne
liquid the number density, although anisotropic, relaxes in a
comparable time scale as that of an isotropic liquid. In the
case of a liquid in the pre-transition region of the I-N transition
the non-ergodicity is due the slow relaxation of the director.
The sharp fall in $\gamma$ shows that mean square enthalpy
($\Delta H$) fluctuations decreases compared to
the same for energy ($\Delta E$) fluctuations. If we write the
enthalpy change at constant pressure as
$\Delta H = \Delta E + P\Delta V$, then the present result suggests
that a positive change in $\Delta V$
(melting of pseudo-nematic domains to the isotropic phase) is
accompanied by a negative change in energy, which is in accordance with the
understanding that the nematic phase has lower internal energy
than the isotropic phase. On the other hand,  the formation of a local
nematic domain due to interaction with the propagating sound wave can
give rise to a negative $\Delta V$ and positive $\Delta E$.
Similarly, increase of sound attenuation is due to the interaction of
the sound propagation with the melting/formation of pseudo-nematic domains which
consumes energy from the sound wave.

\section{\label{ds4}Concluding remarks}
In this paper we have presented  a study of  I-N transition from
the perspective  of  molecular hydrodynamics using molecular
dynamics simulations. In a system of 1125 particles the smallest wavenumber
that can be studied is approximately is $k\sim0.4$ in the range of the
number densities $\rho^*$ = 0.285 to $\rho^*$ = 0.32. The
calculation finds a fairly well defined Brillouin peak in the
range of the wavenumbers studied. These simulations show the role
of density fluctuation and the orientational ordering in development
of anisotropy at intermediate wavenumebrs in the density relaxation near the I-N transition.
However, at large wavenumbers the calculation shows only marginal
evidence for the existence and propagation of the density
fluctuations. The main results of this work  are that it shows
the effect of the formation of the director on the relaxation of
fluctuations of the collective density
of the system.  A fit of intermediate scattering function
to the linearized hydrodynamic model at small wavenumbers
($k=2\pi/L$) show the variation of the thermodynamic as
well as transport properties across I-N transition.
The ratio of the specific heats shows a discontinuity-like feature  near
the I-N transition. The thermal diffusion shows a  dip;
a similar behavior is reported in the recent
experiments of Marinelly {\it et al.}\cite{tdiff}. The sound
dispersion coefficient also shows a divergence like behavior
near I-N transition. The sound velocity calculated
from the peak position if the Brillouin peak deviate from its
monotonic increase near I-N transition also that is in qualitative
agreement with the experimental results of Eden \cite{snd:expt1}
{\it et al.}.

This work demonstrates the effect of spatial and temporal propagation of
anisotropy of the molecular potential in the medium at different state
points across the I-N transition. Also effect of this in the
thermodynamic and transport coefficients. Thus during the I-N transition
the system shows the non-ergodicity due to the formation of
director, that slowly relaxes due to the mutual coupling of
the orientation within a pseudo nematic domain.  Our results
qualitatively agrees with experimental results even though the
potential used is a simple ellipsoidal model. This shows
that these results are related directly the anisotropy of the
inter molecular potential.

In the context of continuing interest of the comparison between the dynamics
of supercooled liquid and liquid crystals\cite{bb:fay1,bb:fay2,bb:ppj1,bb:ppj4},
it is interesting to note that the anomalies observed in sound attenuation
and sound absorption coefficients are not known to exist (at least significantly)
in the former.
This outlines the uniqueness of the liquid crystalline systems where
large pseudo-nematic domains form and melts near the I-N boundary. Such a
physical process does not exist in the supercooled liquid at a thermodynamic
level.

We have argued that most of the features observed can be qualitatively
understood from the melting/formation of the pseudo-nematic domains.
A complete theory of the observed hydrodynamic anomalies need to
consider translation-rotation coupling\cite{bb2:cha,bb2:rw} and is yet to be
developed.

We are not aware of the existence of any theoretical treatment of
the effects of slow down in the decay of the orientational time
correlation function of nematogens near their I-N transition, on the
isotropic dynamic structure factor. In earlier study, we have
reported observation of an anomalous viscoelasticity of the
nematogens near the I-N transition.\cite{bb:ppj2} Both
the problems require inclusion of the effects of orientational slow
down on the isotropic density relaxation. This is clearly a
non-linear problem. While the orientational order parameter S is a
non-conserved order parameter, the isotropic density $\rho$ is a
conserved order parameter. The Landau-de Gennes-Ginzburg free energy
for the isotropic phase of the liquid crystal can be written down
 without much difficulty. However, difficulty lies in the fact that 
the total free energy function should be a functional of both $S(\bf r)$ 
and $\rho(\bf r)$, with a coupling between the two fields. A
molecular hydrodynamic theory along the above line needs to be
developed but still appears to be non-existing.

A valid criticism of the present work is the finite size of the system simulated,
although N=1125 is quite large (and demanding) for long MD simulations,
particularly when orientational degree of freedom is present. Also note that
Gay-Berne is quite a complex potential to study. In addition to N=1125, we
have also studied extensively the size N=500. The basic features remain unchanged. Nevertheless, the
k-->0 limit is expected to be sensitive to the system size, and a system size
dependent study of the present results could be a worthwhile exercise.
The intermediate wavenumber results, however, are not expected to
be sensitive to system size. It will be worthwhile to study this range  by neutron scattering.

\begin{acknowledgments}
This work is supported in parts by grants from DST, India.
\end{acknowledgments}

\end{document}